\documentstyle[prd,aps,preprint,graphics]{revtex}
\begin{document} 
\newcommand{\cL}{{\mathcal{L}}}  
\title{Constraints on Axion Models from $K^+\rightarrow\pi^+ a$} 
\author{Mark Hindmarsh\cite{mhaddress}}
\author{Photis Moulatsiotis\cite{pmaddress}} 
  \address{Centre for Theoretical Physics \\
  University of Sussex \\ Brighton BN1 9QJ \\ U.K} 
\date{November 1998}
\preprint{SUSX-TH-98-017, hep-ph/9807363}
      
\maketitle
\begin{abstract}

We explore a new class of axion models in which some, but not all, of the 
left-handed quarks have a Peccei-Quinn symmetry. These models are potentially 
afflicted by flavour changing neutral currents. We derive the bounds on the 
Peccei-Quinn symmetry-breaking scale from bounds on the $K^+\rightarrow \pi^+ a$ 
branching ratio, showing that in some cases they are even stronger than the 
astrophysical ones, but still not strong enough to kill off the models.

\end{abstract}
\pagebreak

\section{Introduction}

One of the most persistent problems in particle physics is the so called {\it 
strong CP problem}. The CP invariance of the strong 
interactions can, in principle, be spoiled by the addition of the allowed 
`$\theta$-term' \cite{cdgjr}

\begin{equation}
\cL_\theta=\theta{g_{s}^2\over 32\pi^2}G_{b}^{\mu\nu}{\tilde{G}_{b\mu\nu}}.
\end{equation}
However, experimental limits on the neutron dipole moment imply that $\theta < 
10^{-9}$ \cite{edm,edmexp}. This unnatural 
smallness of the $\theta$-parameter is precisely the strong CP problem.

Over the years many solutions have been proposed for resolving this puzzle. A 
most appealing idea is the one brought forward by Peccei \& Quinn \cite{pq}. 
They postulated the existence of an extra global anomalous $U(1)$ symmetry. The 
accomodation of the new charges requires, at least, one extra Higgs doublet. 
With the help of the PQ-symmetry $\theta$ becomes dynamical and is driven to 
zero. 
However, since the new symmetry is not manifest in our 
world, it has to be spontaneously broken. As a result of the Goldstone theorem, 
a pseudoscalar particle appears \cite{ww}, called the {\it axion}. Although one 
would expect the axion to be massless, since it is the Nambu-Goldstone (NG) 
boson of a global 
symmetry, it is actually not. The reason is that, as mentioned before, the 
PQ-symmetry is an anomalous one, spoiled by instanton effects, a fact that 
forces the axion to pick up a small mass through the axion-gluon-gluon anomaly. 

In principle, there is a plethora of axion models as there are many ways of 
assigning the PQ charges to the quark fields, and a lot of freedom to introduce 
extra Higgs fields. The original model \cite{ww}, where 
all quarks were assigned the same charge and where two Higgs doublets were used, 
was experimentally ruled out. In order for the Peccei-Quinn solution to survive, 
another class of axion models was invented, called the invisible axion models, 
where a Higgs singlet was added \cite{ksvz,dfsz}. As a result, the axion decay 
constant $v_a$ became much higher than the electroweak scale and this enabled 
the axion coupling to matter to become much smaller than those of the original 
model. However, the value of $v_a$ cannot be arbitrary. It is constrained 
from below by astrophysical observations and from above by cosmological 
considerations. The current values on 
these limits are $10^{10}<v_a <10^{11-12}\mathrm{GeV}$ \cite{raff,pww,shell} 
leaving a small window for the axion to exist. The value of the upper limit 
depends on the cosmological scenario one favours, that is either inflation or 
cosmic strings, the latter being the most restrictive.

The two mostly talked about axion models are the KSVZ \cite{ksvz} and the DFSZ 
\cite{dfsz}. This does not mean that they are the only ones 
allowed, as we have demonstrated in a recent paper \cite{hm}, in which we 
explored the consequences of assigning different PQ charges to different 
right-handed quarks. In this paper we allow the left-handed quark doublets to 
have different PQ charges, which forces us to confront the problem of flavour 
changing neutral currents (FCNCs). Drawing on the work  of Feng et al 
\cite{feng}, we find that FCNCs constrain the axion scale even more strongly 
than the astrophysical arguments (with one exception).
The most stringent limit comes from the decay $K^+\rightarrow\pi^+ 
a$ \cite{kpa}. There are of course other bounds on the 
axion scale, for example $\mu\rightarrow ea$ \cite{feng}. However, these turn 
out not to be as strong as the $K^+\rightarrow\pi^+ a$ constraints, so we do not 
consider them here. 

The class of 
models we consider represent a minimal digression from the DFSZ model, in that 
instead of all the left-handed quarks having the same PQ charge, only two of 
them do. That requires further extension to the Higgs sector, by adding at least 
one (two in the supersymmetric 
case) new Higgs doublet(s). As a consequence there are up to four other neutral 
scalars and up to three other pseudoscalars in our 
theories, which could contribute to CP-violating processes, in particular 
neutral meson mixing (a fifth scalar coming from the singlet is irrelevant for 
low energy applications). Limits on the mixing translate to limits on the masses 
of 
these pseudoscalars of about a few hundred GeV to a few TeV, which in turn bound 
parameters in a complicated three or four-Higgs potential. Such potentials are 
somewhat problematic in general, as there is a severe hierarchy problem to solve 
in order to separate the electroweak and Peccei-Quinn breaking scales.

\section{Description of the models}

The main idea in this paper is to assign appropriate PQ charges to the Higgs 
fields and consequently to quarks, so that FCNCs can be present at the tree 
level 
of the axion-quark couplings. Then, it will be possible to derive constraints on 
the axion decay constant based on recent experimental data from the absence of 
such processes. In order to take advantage of FCNCs, we give different charges 
to the {\it left}-handed sectors of some of the quarks also (unlike the DFSZ 
where 
all left and all right handed quarks have the same PQ charges). This is a 
special case of a 
minimal change from the DFSZ. The general consistency rules that govern such 
digressions are the topic of future work \cite{hm1}. For constructing these 
models one needs at least three doublets $\phi_n$ and one singlet 
$\phi$. If we allow four Higgs doublets, we have the possibility of making the 
model supersymmetric, as discussed below. The general structure of the Yukawa 
couplings is

\begin{equation}
\cL_Y=f^{n_u}_{ij}(\bar q^\prime_{Li}\phi_{n_u}u^\prime_{Rj})+f^{n_d}_{ij} (\bar 
q^\prime_{Li}\phi_{n_d}d^\prime_{Rj})+h.c
\end{equation}
where $n_u=1,3$, $n_d=2,4$ and $i,j=1,2,3$ are flavour indices. This will result 
in six different models depending on which quarks have PQ charges. Following the 
notation of our previous paper \cite{hm}, in the first 
three the `special' doublets are either the $(u{,}d)_L$, or $(c{,}s)_L$, or 
$(t{,}b)_L$, labeled by I, IV, II, and in the last three ones either 
$(u{,}d)_L$ {\em and} $(c{,}s)_L$, or 
$(u{,}d)_L$ {\em and} $(t{,}b)_L$, or $(c{,}s)_L$ {\em and} $(t{,}b)_L$ 
respectively, 
labeled by V, III, VI.
In the absence of supersymmetry, where the appearence of $\tilde\phi_3$ is 
forbidden as the superpotential must be holomorphic,
one can put $\phi_4=\tilde\phi_3\equiv i\sigma_2\phi_3^*$. Note that all quarks 
are left as flavour eigenstates for the time being. The general lagrangian, part 
of which is the Yukawa sector mentioned above, possesses a PQ symmetry. The most 
general  PQ transformations are 

\begin{eqnarray}
u_{Ri}^\prime & \longrightarrow & e^{i\alpha {T}^{ij}_{Ru}}u_{Rj}^\prime 
\nonumber \\
  d_{Ri}^\prime & \longrightarrow & e^{i\alpha {T}^{ij}_{Rd}}d_{Rj}^\prime 
\nonumber \\
  q_{Li}^\prime & \longrightarrow & e^{i\alpha {T}^{ij}_L}q_{Lj}^\prime \\
  \phi_n & \longrightarrow & e^{iQ_n\alpha}\phi_n, \hspace{.5cm} n=1,2,3,4 .
 \nonumber
\end{eqnarray}
where $Q_1=Q_2=1$ and $Q_3=Q_4=0$. The transformation matrices for the 
left-handed quarks, $T_L$, and the right-handed $u$-type quarks, $T_{Ru}$, are 
listed in Table \ref{t:transmat} for every model. In the case of the 
right-handed $d$-type quarks the transformation matrices $T_{Rd}=0$ in all cases 
thus not listed in the Table.  As an example, let us consider 
Model I, the tranformation matrices of which are
\begin{eqnarray}
T_L= 
\left( \begin{array}{ccc}  
1 & 0 & 0 \\ 0 & 0 & 0 \\ 0 & 0 & 0
  \end{array} \right),  \qquad
T_{Ru}=\left( \begin{array}{rcc}  
-1 & 0 & 0 \\ 0 & 0 & 0 \\ 0 & 0 & 0
  \end{array} \right), \qquad T_{Rd}=0. \nonumber  
\end{eqnarray}
Their application fixes the Yukawa couplings $f^{n_{u,d}}_{ij}$ to have zeros in 
certain entries. For the $u$-type quarks the relevant Yukawa matrices are 
$f^1_{ij}$ and $f^3_{ij}$. In the first one, the only non-zero element is 
$f^1_{11}$ and in the second we must have $f^3_{i1} = f^3_{1j} = 0$. On the 
other hand, for the $d$-type quarks, $f^2_{ij}$ and $f^4_{ij}$ being the 
appropriate Yukawa matrices, we need $f^2_{2j}=f^2_{3j}=0$, and $f^4_{1j}=0$.

Our next step is to determine the axion decay constant in terms of the vacuum 
expectation values of the Higgs fields and the mixing with the $Z^0$. Suppose 
that $a^\prime$ and $Z$ are the would-be Goldstone bosons {\it before} 
instantons are taken into account. We define $a^\prime$ to be the massless 
axion and $Z$ the longitudinal degree of freedom of the $Z^0$ boson. Suppose 
also that $\alpha$ and $\alpha_Z$ are the angles conjugate to the PQ and Z 
transformations, so that

\begin{eqnarray}
  \phi_1 &=& {1\over\sqrt{2}}
  \left( \begin{array}{c}
  v_1 \\ 0 \end{array} \right) \exp{[i(2\alpha-\alpha_Z)]}, \hspace{0.5cm}
  \phi_2={1\over\sqrt{2}}
  \left( \begin{array}{c}
  0 \\ v_2 \end{array} \right) \exp{[i(\alpha+\alpha_Z)]},        
   \nonumber \\
  \phi_3 &=& {1\over\sqrt{2}}
  \left( \begin{array}{c}
  v_3 \\ 0 \end{array} \right) \exp{(-i\alpha_Z)}, \hspace{1.19cm}
  \phi_4={1\over\sqrt{2}}
  \left( \begin{array}{c}
  0 \\ v_4 \end{array} \right) \exp{(i\alpha_Z)}, \\
  \phi&=&{1\over\sqrt{2}}v\exp{(-i\frac{3}{2}\alpha)}, \nonumber
\end{eqnarray}
where $v_1$, $v_2$, $v_3$, $v_4$, $v$ are the vacuum expectation values of the 
Higgs fields. In order to separate the axion from the $Z^0$, one comes to the 
following equation

\begin{equation}
  \left( \begin{array}{c}a^\prime \\ Z \end{array}\right)=\left(   
  \begin{array}{lr} v_a & 0 \\ v_{10} & v_{11} \end{array}\right)\left(  
  \begin{array}{c}\alpha \\ \alpha_Z \end{array}\right)
  \label{e:mixmat}
\end{equation}
where the $2\times 2$ matrix on the right hand side of (\ref{e:mixmat}) is the 
most general one compatible with this requirement. A simple comparison of 
(\ref{e:mixmat}) with the kinetic terms of the NG bosons (coming from the 
kinetic terms of the Higgs fields) \cite{hm} gives the expression 
for the axion decay constant and for the electroweak breaking scale

\begin{eqnarray}
  v_a &=& \sqrt{\frac{9}{4}v^2+{{(4v_1^2+v_2^2)(v_3^2+v_4^2)+9v_1^2v_2^2}\over 
{v_1^2+v_2^2+v_3^2+v_4^2}}}, \\
  v_{EW} &=& v_{11}=\sqrt{v_1^2+v_2^2+v_3^2+v_4^2}=\mathrm {246 \hspace{.2cm} 
GeV}, \\
  v_{10} &=& {v_2^2-2v_1^2\over{\sqrt{v_1^2+v_2^2+v_3^2+v_4^2}}}.
\end{eqnarray}
As we see, it has the same value for all six models. Furthermore, it is 
essentially equal to $v$, the PQ symmetry breaking scale.

\section{Axion-quark couplings and induced FCNCs}

Let us take now a closer look at the axion couplings to quarks. In the flavour 
basis the relevant term of the QCD lagrangian is
\begin{equation}
  \cL_{int}=-{\partial^\mu a^\prime\over 2v_a}[\bar u'_i\gamma_\mu ((1-\gamma_5) 
T^{ij}_L +(1+\gamma_5)T^{ij}_{Ru})u'_j+\bar d'_i\gamma_\mu (1-\gamma_5) T^{ij}_L 
d'_j]
  \label{e:Lintfl}
\end{equation}
It is possible to diagonalise, the generally non-diagonal, quark mass 
matrix by a bi-unitary transformation. In more detail, there are unitary 
transformations that relate the flavour basis with the mass one, of each of 
the $u$ and $d$-type quarks
\begin{eqnarray}
u'_{Ri}=U^{ij}_R u_{Rj}, \qquad d'_{Ri}=D^{ij}_R d_{Rj},  \nonumber \\ 
u'_{Li}=U^{ij}_L u_{Lj}, \qquad d'_{Li}=D^{ij}_L d_{Lj}.  \nonumber
\end{eqnarray}
Applying these transformations to (\ref {e:Lintfl}) and going to the mass 
basis, the lagrangian takes the form
\begin{equation}
  \cL_{int}={\partial^\mu a^\prime\over 2v_a}[2\bar u_i\gamma_\mu \gamma_5 
S^{ij}_{Lu} u_j-\bar d_i\gamma_\mu (1-\gamma_5) S^{ij}_{Ld} d_j]
  \label{e:Lintm}
\end{equation}
where $S_{Lu}=U^\dag_L T_LU_L$ and $S_{Ld}=D^\dag_L T_LD_L$. By definition, the 
Cabbibo-Kobayashi-Maskawa matrix is $V_{CKM}=U^\dag_L D_L$, so it is obvious 
that
\begin{eqnarray}
S_{Ld}=V_{CKM}^\dag S_{Lu} V_{CKM},
\label{e:SV}
\end{eqnarray}
thus being possible for FCNCs to be present in the $d$-type quark sector, both 
in the vector and in the axial-vector part of the Lagrangian. Furthermore, it is 
evident from the 
structure of the $u$-type Yukawa couplings, that $U_L$ and $U_R$ have a block 
diagonal form and thus $S_{Lu}=T_L$ in all cases. So eq. (\ref{e:SV}) becomes
\begin{eqnarray}
S_{Ld}=V_{CKM}^\dag T_L V_{CKM},
\label{e:TV}
\end{eqnarray}
Combining the data from Table \ref{t:transmat} and eq. (\ref{e:TV}) one finds
\begin{eqnarray}
T_d=\left( \begin{array}{ccc}  
V_{u_id}^* V_{u_id} \hspace{.3cm} & V_{u_id}^* V_{u_is} \hspace{.3cm} & 
V_{u_id}^* V_{u_ib} 
\\V_{u_is}^* V_{u_id} \hspace{.3cm} & 
V_{u_is}^* V_{u_is} \hspace{.3cm} & V_{u_is}^* V_{u_ib} \\V_{u_ib}^* V_{u_id} 
\hspace{.3cm} 
& V_{u_ib}^* V_{u_is} \hspace{.3cm} & V_{u_ib}^* V_{u_ib} 
  \end{array} \right)
\label{e:m124}
\end{eqnarray}
for Models I, II, IV, where $i=1,2,3$ labels the PQ-charged quarks and
\begin{eqnarray}
T_d=\left( \begin{array}{lcr}  
V_{u_id}^* V_{u_id}+V_{u_jd}^* V_{u_jd} \hspace{.4cm}& V_{u_id}^* 
V_{u_is}+V_{u_jd}^* V_{u_js} \hspace{.4cm}& V_{u_id}^* V_{u_ib}+V_{u_jd}^* 
V_{u_jb} \\ V_{u_is}^* V_{u_id}+V_{u_js}^* V_{u_jd} \hspace{.4cm}& 
V_{u_is}^* V_{u_is}+V_{u_js}^* V_{u_js} \hspace{.4cm}& V_{u_is}^* 
V_{u_ib}+V_{u_js}^* V_{u_jb} \\ V_{u_ib}^* V_{u_id}+V_{u_jb}^* V_{u_jd} 
\hspace{.4cm}& V_{u_ib}^* V_{u_is}+V_{u_jb}^* 
V_{u_js} \hspace{.4cm}& V_{u_ib}^* V_{u_ib}+V_{u_jb}^* V_{u_jb} 
  \end{array} \right)
\label{e:m356}
\end{eqnarray}
for Models III, V and VI, where $i\neq j$ also labeling the relevant PQ-charged 
quarks. As we shall see in 
the following section the interesting part of 
the interaction lagrangian (\ref{e:Lintm}) is the one giving the transition 
of $s\rightarrow d$ quarks. In this case 
\begin{equation}
  \cL_{int}=-{\partial^\mu a'\over 2v_a}
[\bar s\gamma_\mu (g_{sd}^V+g_{sd}^A\gamma_5) d+h.c]
  \label{e:Lintsd}        
\end{equation}
$g_{sd}^V$ and $g_{sd}^A$ being, by definition, the vector and axial vector 
parts of the $a-s-d$ coupling. 
Combining eqs. (\ref{e:Lintm}), (\ref{e:m124}), (\ref{e:m356}) and 
(\ref{e:Lintsd}) one finds
\begin{eqnarray}
g_{sd}^V=\left\{ \begin{array}{ll}
  V_{u_id}^* V_{u_is} & {\textrm{Models I, II, \& IV}} \\
  V_{u_id}^* V_{u_is}+V_{u_jd}^* V_{u_js} \hspace{1cm} & {\textrm{Models III, V, 
\& VI}} 
  \end{array} \right.
 \label{e:gvsd}
\end{eqnarray}
Concerning the axial coupling, as we shall see in the next section  is of no 
importance, since only the vectorial one is involved in the calculation of the 
rate $K^+\rightarrow\pi^+ a$. The values for the CKM elements used are \cite{cp} 
$|V_{ud}|\approx 0.98$, $|V_{us}|\approx 0.22$, $|V_{cd}|\approx 0.22$, 
$|V_{cs}|\approx 0.97$, $|V_{td}|\approx 9.1\times 10^{-3}$ and $|V_{ts}|\approx 
3.9\times 10^{-2}$. 

\section{Experimental constraints}

As described in the previous section, the axion can take part in 
flavour-changing 
processes. One can extract lower bounds on the axion decay constant from 
experimental data concerning such processes. The tighter constraints come from 
transitions between the first two generations, whereas bounds involving the 
third one are much weaker. The processes that produce the most stringent limit  
are the ones involving rare $K$ decays. The most suitable one
for our discussion is $K^+\longrightarrow \pi^+ a$, the decay rate of which is 
\cite{feng}
\begin{equation}
\Gamma (K^+\rightarrow \pi^+ a) = 
\frac{1}{16\pi}\frac{m_K^3}{v_a^2}{g_{sd}^V}^2(1-\frac{m_{\pi}^2}{m_K^2})^3 
|F_1(0)|^2
\end{equation}
where $F_1(0)$ is the form factor 
$F_1(q^2)(p+p^{'})^{\mu}=\langle\pi^+(p^{'})|\bar s \gamma^{\mu}d|K^+(p)\rangle$ 
at zero momentum transfer and is of order unity, being exactly 1 in the case of 
exact $SU(3)$ flavour symmetry. The experimental data sets an upper limit on the 
branching ratio \cite{kpa} $\mathrm{Br}(K^+\rightarrow \pi^+ a)<3.0\times 
10^{-10} $ (at 90\% confidence). This leads to a lower bound on the axion energy 
scale
\begin{equation}
v_a>1.7\times 10^{11}\times g_{ds}^V \hspace{1cm}\mathrm{GeV}
\end{equation} 
Taking into account eq. (\ref{e:gvsd}) and the current values for the relevant 
CKM matrix elements \cite{cp}, the above expression yields lower limits on 
$v_a$. The results are listed in Table \ref{t:DevAxLims}.

\begin{figure}[ht]
  \centering
  \scalebox {0.75} {\includegraphics {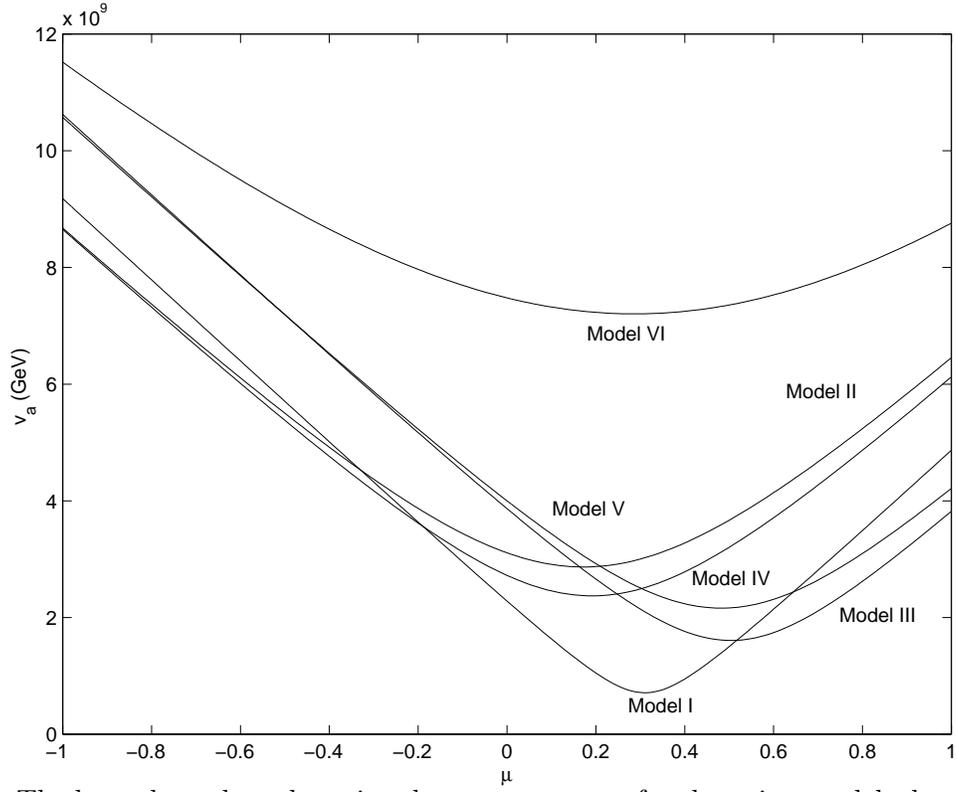}}
  
  \caption {\label{f:astr}The lower bound on the axion decay constant $v_a$ 
for the axion models described in the text, plotted as a function of 
$\mu=\frac{v_2^2-v_1^2}{v_{EW}^2}$.}
  \end{figure}

It will be very instructive to compare these results with the astrophysical ones 
since the latter are so far considered to be the most severe. It is a well known 
fact that among the astrophysical limits the far more restrictive are the ones  
coming from SN1987A, bounding the axion-nucleon-nucleon coupling. The limit of 
this constraint, taking many body effects into account, is \cite{1987}

\begin{equation}
(h_{ap}^2+2h_{an}^2)^{1/2}<2.85\times 10^{-10}.
\end{equation}
A similar analysis as the one performed in \cite{hm}, normalising the PQ charges 
for $N=1$ or $N=2$ depending on the case, yields the following lower 
bounds on the axion decay constant

\begin{equation}
v_a>0.35\times 10^{10}\times (A\mu^2-B\mu+C)^{1/2}\mathrm{GeV}
\label{e:astrbound}
\end{equation}
where 
$$\mu\equiv\frac{v_{10}}{v_{11}}=\frac{v_2^2-v_1^2}{v_{EW}^2}$$
and $\frac{-v_1^2}{v_{EW}^2}\leq\mu\leq\frac{v_2^2}{v_{EW}^2}$. The values of 
the coefficients $A$, $B$ and $C$ are summarised in Table \ref{t:astro} for each 
model. 
One can easily see from eq. (\ref{e:astrbound}) that the most stringent limits 
come for $\mu=-1$, in the limit where $v_2=v_3=v_4=0$. However, comparison of 
these results (plotted in Fig. \ref{f:astr}) to the ones coming from FCNCs 
reveals that for all models, except the third one, the latter constraints are 
more severe. 
Especially in the case of the fifth model the limit is almost up to $10^{11}$ 
GeV, very close to the upper bound coming from the cosmological scenario of 
cosmic strings \cite{shell}. The exception of the third model is due to the fact 
that $V_{td}^* V_{ts}\ll V_{ud}^* V_{us}\approx V_{cd}^* V_{cs}$, 
also responsible  for the similar behavior of the rest of them.

\section{Constraints on massive scalars and pseudoscalars}

A theory with four Higgs doublets and a Higgs singlet has an additional five  
massive neutral scalars  and three massive pseudoscalars. Apart from the  scalar 
coming from the singlet, the rest can in principle mediate neutral meson mixing 
in the low energy sector. For example, $B\bar B$ mixing proceeds via the 
operators
\begin{eqnarray}
O_{B\bar B}^S&=&h^k_{bd}(\bar d\: b)\frac{1}{M^2_{h_k}}(\bar b\: d), \nonumber 
\\
O_{B\bar B}^P&=&h^k_{bd}(\bar d\gamma_5 b)\frac{1}{M^2_{A_k}}(\bar b\gamma_5 d),
\label{e:scapseuop}
\end{eqnarray}
for the scalar and the pseudoscalar case respectively, where $h^k_{bd}$ is the 
flavour-changing coupling and $M_{h_k},M_{A_k}$ are the masses of the $k-$th 
scalar and  pseudoscalar. Evaluating the matrix elements for $B\bar B$ mixing in 
each case one finds \cite{atwetal}
\begin{eqnarray}
\langle B^0|O_{B\bar B}^S|\bar B^0\rangle&=&\frac{1}{6}f^2_B m^2_B \left( 
\frac{m^2_B}{(m_b+m_d)^2}+1\right), \nonumber \\
\langle B^0|O_{B\bar B}^P|\bar B^0\rangle&=&\frac{1}{6}f^2_B m^2_B \left( 11 
\frac{m^2_B}{(m_b+m_d)^2}+1\right)
\label{e:matrelem}
\end{eqnarray}
where $f_B$ and $m_B$ are the $B$ decay constant and mass respectively and 
$m_b,m_d$ are the masses of the $b$ and $d$ quarks. These expressions were 
evaluated using the vacuum insertion approximation in both cases, in order for a 
direct comparison to be possible.  Of course one may argue that vacuum 
insertion, 
although a good approximation in the pseudoscalar case, is not as good in the 
scalar. For an order-of-magnitude estimate it should however suffice. In 
the heavy quark approximation, where $m_B\simeq m_b+m_d$, this gives a mass 
splitting of \cite{atwetal}
\begin{eqnarray}
\Delta m_B&\simeq&\frac{2}{3}h^2_{bd}\frac{f^2_B}{M^2_h}m_B, \nonumber \\
\Delta m_B&\simeq&4h^2_{bd}\frac{f^2_B}{M^2_A}m_B,
\label{e:masssplit}
\end{eqnarray}
for the scalar and the pseudoscalar interactions respectively, where we have 
dropped the $k$ index, since we assume that the larger mass splitting is due to 
the lightest among the scalars and pseudoscalars. It is obvious that 
pseudoscalars provide stronger constraints by a factor of 6. In order to extract 
constraints for their masses we must first comment on the values $h_{bd}$ can 
take. Cheng \& Sher \cite{chsh} have argued that in a wide class of models 
where no fine-tuning is assumed, the flavour-changing couplings are of the order 
of the geometric mean of the Yukawa couplings of the generations involved, 
$h^2_{ij}=\kappa_{ij}^2 f_i f_j$. Thus in our case we have
\begin{equation}
h^2_{bd}=\kappa_{bd}^2 f_b f_d=\kappa_{bd}^2\frac{m_b 
m_d}{\langle\phi_n\rangle\langle\phi_m\rangle}
\label{e:f-ccoupl}
\end{equation}
where $n$ and $m$ take the values 2 or 4 depending on the model. Since we do not 
know the expectation value of each Higgs field we will attempt an estimate of 
the order-of-magnitude for the Higgs masses. Taking the value for the decay 
constant of the $B$ meson to be $f_B\approx 170\;\mathrm{MeV}$ from lattice 
calculations \cite{flysachr} and the experimental value for $\Delta m_B<3\times 
10^{-13}\;\mathrm{GeV}$ \cite{pdg98}, we obtain
\begin{equation}
M_h>\kappa_{bd}\times 1.3\;{\mathrm{TeV}}, \qquad M_A>\kappa_{bd}\times 3.2 \; 
{\mathrm{TeV}}.
\label{e:massconstr}
\end{equation}
As we can see the results are in the TeV scale.  This means that it is 
necessary only to make the not-so-unreasonable assumption that $\kappa$'s 
which involve the first generation can take values much less than one, in order 
that the constraints (\ref{e:massconstr}) can be further weakened and the masses 
can take values within the range of a few hundred GeV, a natural Higgs mass for 
an electroweak theory.  However, we are still left with a hierarchy problem,
that is common to all invisible axion models; how to account for the wide range
of scale between 100 GeV and $10^{12}$ GeV. Our model is no improvement
is this regard.  It should in principle be possible to make the models 
supersymmetric, when there are four Higgs doublets, in which case it is 
consistent to set the singlet-doublet couplings to be extremely small without 
fear of radiative corrections.

\section{Conclusions}

It is evident from the above analysis that there is a lot of freedom in choosing 
the PQ charges in the quark sector. In this paper we studied a new class 
of axion models, where the left-hand sector of certain quark flavours (but not 
all) were assigned PQ charges. As a consequence, FCNCs are induced, which can be 
used to provide a lower bound on the axion decay constant. It was shown that for 
certain models the limits on these bounds are more severe than those coming from 
astrophysics, with the most striking example the case of Model V, although not 
severe enough to rule them out. We have also estimated constraints coming from 
neutral meson mixing induced by other scalars and pseudoscalars, which constrain 
their masses to be in the range of a few hundred GeV to a few TeV. 

\acknowledgments

We wish to thank H.R. Quinn and B. de Carlos for useful discussions. 
M.H. is supported by PPARC Advanced Fellowship B/93/AF/1642 and by PPARC grant 
GR/K55967.

\begin{table}
\begin{tabular}{|l|lll|}
Model & $T_L$ & $T_{Ru}$ & \\
\hline
I & diag $(1,0,0)$ & diag $(-1,0,0)$ & \\
II & diag $(0,0,1)$ & diag $(0,0,-1)$ & \\
III & diag $(1,0,1)$ & diag $(-1,0,-1)$ & \\
IV & diag $(0,1,0)$ & diag $(0,-1,0)$ & \\
V & diag $(1,1,0)$ & diag $(-1,-1,0)$ & \\
VI & diag $(0,1,1)$ & diag $(0,-1,-1)$ &
\end{tabular}
 \caption{\label{t:transmat}Transformation matrices for all left-handed and 
$u$-type right-handed quarks, concerning the models discussed in the text. 
$T_{Rd}=0$ in all cases thus not listed.}
\end{table}

\begin{table} 
\begin{tabular}{|l|llll|}
Model & Charged & Vector & Axion scale &\\
 & doublets & coupling & ($\times 10^{10}$ GeV) &\\
\hline
I & $ud$ & $V_{ud}^* V_{us}$ & $3.7$ &\\
II & $tb$ & $V_{td}^* V_{ts}$ & $0.0061$ &\\
III & $ud,tb$ & $V_{ud}^* V_{us}+V_{td}^* V_{ts}$ & $3.7$ &\\
IV & $cs$ & $V_{cd}^* V_{cs}$ & $3.6$ &\\
V & $cs,ud$ & $V_{ud}^* V_{us}+V_{cd}^* V_{cs}$ & $7.3$ &\\
VI & $cs,tb$ & $V_{cd}^* V_{cs}+V_{td}^* V_{ts}$ & $3.6$ &
\end{tabular}
\caption{\label{t:DevAxLims}Limits on the axion scale $v_a$ from 
the flavour-changing process $K^+\to \pi^+ a$.}
\end{table}

\begin{table}
\begin{tabular}{|l|cccc|}
Model & $A$ & $B$ & $C$ &\\
\hline
I & $3.98$ & $2.47$ & $0.42$ &\\
II& $3.98$ & $1.37$ & $0.79$ &\\
III&$3.98$ & $4.01$ & $1.22$ &\\
IV& $3.98$ & $1.52$ & $0.61$ &\\
V&  $3.98$ & $3.84$ & $1.31$ &\\
VI& $3.98$ & $2.29$ & $4.57$ &
\end{tabular}
\caption{\label{t:astro} Table of the values of the coefficients in Eq. 
(\ref{e:astrbound}) for the axion models considered in the text.}
\end{table}

\end{document}